\documentclass[12pt,aps,floats]{revtex4}
\usepackage{graphicx}
\usepackage{epsfig}

\newcommand{\be}{\begin{equation}}
\newcommand{\ee}{\end{equation}}
\newcommand{\bea}{\begin{eqnarray}}
\newcommand{\eea}{\end{eqnarray}}

\def\simlt{\stackrel{<}{{}_\sim}}

\begin{document}
\title{Acoustic peaks in the CMB: a matter of standard causal 
boundary conditions on primordial density anisotropies}

\author{David H. Oaknin}
--
\affiliation{
Department of Physics, Technion, Haifa, 32000, Israel. \\
e-mail: doaknin@physics.technion.ac.il}

\begin{abstract}
The pattern of acoustic peaks in the sub-horizon power spectrum of primordial 
density anisotropies at recombination can be naturally understood in the 
framework of standard Friedmann-Robertson-Walker cosmology (without inflation) 
as a consequence of the boundary conditions imposed by the causal horizon on 
the statistical two-points correlation functions: the sub-horizon spectrum is 
discrete (harmonic), with comoving modes located at $k_n = n \frac{\pi}{H^{-1}_{eq}}$, 
$n = 1,2,...$, because the causally connected patch of the universe at 
recombination is compact, with comoving radius $H^{-1}_{eq}$.
The results presented in this paper complement those presented in [1], 
where it was shown that the scale invariance of the primordial density 
anisotropies over comoving scales of cosmological size is also a consequence 
of the boundary conditions imposed by causality. Together these results
lay an appealing theoretical alternative to the inflationary paradigm as the 
ultimate answer for the origin of cosmological structures in standard cosmology.
\end{abstract}

\maketitle

The extremely uniform blackbody radiation at $T \sim 2.7$~K that 
permeates the whole visible universe \cite{Penzias} is a relic signature of the hotter 
and denser earlier stage at $T_{eq} \sim 3 \times 10^3$~K when free 
electrons and protons recombined into atomic hydrogen and the electromagnetic 
radiation decoupled from the thermal cosmic plasma \cite{Peebles}. 
The anisotropies observed over the sky $\frac{\Delta T(\theta,\phi)}{T} \simlt 
10^{-5}$ in the local temperature of this blackbody radiation 
\cite{Spergel:2003cb}, known commonly as the cosmic microwave background~(CMB),
are the imprint left by tiny random primordial density 
fluctuations in the, otherwise, homogeneous and isotropic cosmic plasma at 
the instant of recombination. 

Precision measurements of the tiny anisotropies in the temeperature of the CMB 
over the sky have become an extremely powerful tool in today cosmology. 
Yet the theoretical understanding of the ultimate mechanism that seeded the
primordial density anisotropies at the instant of recombination is unclear. 
It is widely believed that the statistical correlations of these primordial 
anisotropies cannot been explained by causal physics in the framework of the, 
otherwise, extraordinarily successful standard Friedmann-Robertson-Walker~(FRW) 
cosmology. The main reason for this belief is, in vague words, that primordial 
anisotropies extend over comoving volumes of cosmological size, which are much 
larger than the comoving causal horizon at the instant of decoupling 
$H^{-1}_{eq}$. The puzzling apparent contradiction is known in the literature 
as the {\it origin of structures problem} of standard 
cosmology. It is closely related to the so-called {\it horizon problem} and has 
become the source of many speculative ideas in cosmology for, at least, 
the last twenty years. 

The most outstanding of the proposals that have been considered to solve the 
{\it origin of structures problem} and {\it horizon problem} of standard 
cosmology is the so-called inflationary paradigm. 
The paradigm assumes that the patch of the universe visible at present was once 
causally connected in the remote past, at $T \simlt T_{Planck} \sim 10^{28}$~eV, and 
primordial density anisotropies were imprinted then. A finite period of 
exponential expansion, the inflation, would have supposedly stretched that 
diminute comoving patch, beyond the causal horizon, to the size of the 
currently visible universe. The inflation is regarded today as the natural 
solution to the problems of standard cosmology, even though all the attempts 
to implement it in models of particle cosmology have failed.   

In reference \cite{oaknin1} we showed that the analysis that led to formulate 
the {\it origin of structures problem} and {\it horizon problem} of standard 
cosmology were based on an unjustified, and erroneous, approximation 
that had gone unnoticed. We indeed showed that the generic 
statistical features of the primordial density anisotropies over 
comoving scales of cosmological size can be naturally and simply 
explained in the framework of standard FRW cosmology, without any need of an 
earlier stage of inflation in the remote past.
The aim of this paper is to complement the results of ref. \cite{oaknin1} and 
show that also the generic statistical features of the primordial density 
anisotropies over sub-horizon comoving scales can be naturally and simply 
understood in the standard FRW cosmology.

In short: we claim that in standard FRW cosmology (without inflation)
random density anisotropies causally generated at the same instant of 
recombination have the generic statistical features of the 
observed primordial anisotropies. Moreover, these characteristic features 
are indeed derived from the constrain that primordial fluctuations cannot be 
correlated over comoving scales longer than the causal horizon at 
recombination. 

Namely, the referred generic statistical features of the observed primordial 
anisotropies are:

a) scale invariance over comoving length scales $L \gg H^{-1}_{eq}$ of 
cosmological size;

b) a pattern of acoustic (harmonic) peaks over sub-horizon comoving length scales 
$L \simlt H^{-1}_{eq}$, located at comoving momenta $k \simeq n \frac{\pi} 
{H^{-1}_{eq}}$ with $n=1,2,...$ a natural number. \\

In \cite{oaknin1} we addressed feature a). We proved there that random density 
fluctuations causally generated at the instant of recombination, and with 
comoving wavelength $\lambda \simlt H^{-1}_{eq}$ shorter than the comoving 
causal horizon at that instant, produce scale invariant mass anisotropies 
over comoving volumes of cosmological size $L \gg H^{-1}_{eq}$. Scale 
invariance means that the variance of random anisotropies in the extensive
magnitude $M_V$ (the mass/energy contained in a comoving volume $V$)  
is $(\Delta M_V)^2 \sim S$ linearly proportional to the area $S \sim L^2$ of 
the surface that bounds the region of integration, rather than to its 
volume $V \sim L^3$. This feature can be intuitively understood in the
framework of standard FRW cosmology by noticing that, under standard causal 
constrains, only random density fluctuations happening at the boundary surface 
can contribute to fluctuations of the extensive magnitude $M_V$ over 
cosmologically large comoving volumes. We refer the reader to \cite{oaknin1} 
and \cite{oaknin2} for a more detailed explanation of these arguments.  

In this paper we address feature b). We start noticing that in standard FRW 
cosmology it can be consistently (and naturally) claim that the
causal horizon $H^{-1}_{eq}$ at the instant of recombination forces 
the two-point correlation function 
$f({\vec x} - {\vec y}) = \langle \rho({\vec x}) \rho({\vec y})\rangle - 
\langle \rho({\vec x})\rangle \langle \rho({\vec y})\rangle$, ${\vec x}, 
{\vec y} \in {\bf R}^3$, of the homogeneous and isotropic stochastic 
density field $\rho({\vec x})$ to cancel over longer comoving distances,

\begin{equation}
\label{constrain}
f({\vec x} - {\vec y}) = 0, 
\hspace{0.3in} if \hspace{0.2in} 
|{\vec x} - {\vec y}| \ge H^{-1}_{eq}.
\end{equation}
That is, in standard FRW cosmology causality constrains the two-points 
correlation function to a compact support, the comoving sphere of 
radius $H^{-1}_{eq}$. This condition (\ref{constrain}) indeed defines 
$H^{-1}_{eq}$: it is the comoving sound causal horizon at recombination.
If the two points correlation function has 
compact support, it accepts a discrete Fourier transform

\begin{equation}
\label{Fourierdiscrete}
f(|{\vec x} - {\vec y}|) \simeq \left(\frac{\rho^2_0}{2\pi^2}\right)
\sum_{n \in {\bf N}} \left(\frac{\pi}
{H^{-1}_{eq}}\right)^3 \cdot \left(n^2 {\cal P}_n\right)
\frac{sin\left(\frac{\pi}{H^{-1}_{eq}} n \cdot r\right)}
{\left(\frac{\pi}{H^{-1}_{eq}} n \cdot r\right)},
\end{equation}   
with comoving modes located at $k_n = n \frac{\pi}{H^{-1}_{eq}}$ 
with $n = 1,2,...$. This discrete power spectrum reproduce the localization 
of harmonic acoustic peaks in the observable sub-horizon spectrum of 
primordial cosmological density anisotropies, with a first peak located
at $k_1 = \frac{\pi}{H^{-1}_{eq}}$. 

In the theoretical framework we have laid the localization of the acoustic 
peaks is determined by condition (\ref{constrain}). These arguments prove 
that inflation is not necessary to understand the statistical features of
the primordial density anisotropies imprinted in the CMB, either their
scale invariance over cosmologically large comoving volumes nor the pattern
and localization of acoustic peaks over the sub-horizon scales.
Furthermore, if our interpretaion of the origin of primordial structures
is correct it can be used to constrain non-standard cosmologies.
  
As a last brief comment we wish to notice that the acoustic peaks 
in the observable power spectrum of temperature anisotropies in the cosmic 
background radiation are not expected to be sharply located, as expansion
(\ref{Fourierdiscrete}) could induce to think. They should appear as broad
peaks with a width determined by causal effects. 
In this paper, nevertheless, we will not attempt precise calculations of 
such effects, because our aim, for the moment, is simply to show that the 
generic features of the primordial density anisotropies at the instant 
of recombination clearly correspond to those that should be expected if the 
anisotropies were causally generated in the framework of standard FRW cosmology, 
without inflation. In fact, we have shown that the existance in 
standard FRW cosmology of a comoving causal horizon at the instant 
of recombination much shorter than all comoving cosmological length scales 
is not an obstacle for explaining the generic statistical 
features of the primordial density anisotropies imprinted in the CMB. 
Just the contrary, the existance of such horizon is the cause of these 
characteristic features. 

\end{document}